\documentclass[twocolumn,aps,amsmath,amssymb,pre]{revtex4}
\usepackage{graphicx}
\usepackage{epsfig}
\bibstyle{apsrev.bib}

\begin{document}
\input epsf.sty

\title{Weakly coupled, antiparallel, totally asymmetric simple exclusion processes}

\author{R\'obert Juh\'asz}
 \email{juhasz@szfki.hu} 
\affiliation{Research Institute for Solid
State Physics and Optics, H-1525 Budapest, P.O.Box 49, Hungary}

\date{\today}

\begin{abstract}
We study a system composed of two parallel totally asymmetric 
simple exclusion processes with open boundaries, 
where the particles move in the two
lanes in opposite directions and are allowed to jump to the other lane 
with rates
inversely proportional to the length of the system.
Stationary density profiles are determined and the phase diagram of
the model is constructed in the hydrodynamic limit, by 
solving the differential equations describing the steady state of the system, 
analytically for vanishing total current and
numerically for nonzero total current. 
The system possesses phases with a localized shock in the density
profile in
one of the lanes, similarly to exclusion processes endowed with
nonconserving kinetics in the bulk. Besides, the system undergoes a
discontinuous phase transition,
where coherently moving delocalized shocks emerge in both lanes 
and the fluctuation of the global density is 
described by an unbiased random walk. This phenomenon 
is analogous to the phase coexistence observed at the coexistence line of the totally asymmetric simple 
exclusion process, however, as a consequence of
the interaction between lanes, the density profiles are deformed and 
in the case of asymmetric lane change, the motion of the shocks is
confined to a limited domain. 

\end{abstract}

\maketitle

\newcommand{\bc}{\begin{center}}
\newcommand{\ec}{\end{center}}
\newcommand{\be}{\begin{equation}}
\newcommand{\ee}{\end{equation}}
\newcommand{\beqn}{\begin{eqnarray}}
\newcommand{\eeqn}{\end{eqnarray}}

\vskip 2cm
\section{Introduction}

The investigation of interacting stochastic driven diffusive systems
plays an important role in the understanding of nonequilibrium
steady states \cite{liggett,zia}. 
As opposed to equilibrium statistical mechanics, 
phase transitions may occur in these systems even in one spatial
dimension \cite{evans}. 
The paradigmatic model of driven lattice gases is 
the one-dimensional totally asymmetric
simple exclusion process (TASEP) \cite{mcdonald,schutzreview},
 which exhibits boundary induced
phase transitions \cite{krug} and the steady state of which is exactly
known \cite{derrida,schutzdomany}. 
Beside theoretical interest, this model and its numerous variants have
 found a wide range of applications,
 such as the description of vehicular traffic \cite{santen} or modeling
 of transport processes in biological systems \cite{schad}.   
Inspired by the traffic of cytoskeletal motors \cite{howard}, such models were
introduced where a totally asymmetric exclusion
process is coupled to a finite compartment where the motion of
particles is diffusive \cite{nieuwenhuizen,klumpp,muller}.
Recently, the attention has turned to exclusion processes endowed with
various types of reactions which violate the conservation
 of particles in the bulk
 \cite{challet,frey,schutz03,ejs,rakos1,js,klumpp2,levine,ehk,rakos2,pierobon,mobilia,hinsch,oshanin}.   
The simplest one among these models is the TASEP with ``Langmuir
kinetics'', where particles are created and annihilated  also 
at the bulk sites of the
system \cite{frey}---a process, which may serve as a simplified model
for the cooperative motion of molecular motors along a filament from
 which motors can detach and attach to it again. 
For these types of systems, the time scale of nonconserving processes compared
to that of directed motion and the processes at the boundaries is crucial. 
If the nonconserving reactions occur with rates of larger order than
the inverse of the system size $L$, then in the large $L$ limit, 
they dominate the stationary state. 
On the contrary, when they are of smaller order than
$\mathcal{O}(1/L)$, 
they are irrelevant and the stationary state is identical to that of
the underlying driven diffusive system. 
However, in the marginal case when the rates of nonconserving processes
 are of order 
$\mathcal{O}(1/L)$, the 
interplay between them and the boundary processes may result in
intriguing phenomena, such as ergodicity breaking \cite{rakos1,rakos2}  or
the appearance of a localized shock in the density
profile \cite{frey}, which is in contrast to the delocalized shock 
dynamics at the coexistence line of the TASEP
 \cite{schutzdomany,schutz98}. The formation of domain walls 
 can be observed also experimentally in the transport of kinesin
motors in accordance with theoretical predictions \cite{konzack,nishinari,greulich}.

Other systems which have an intermediate complexity compared to exclusion
processes with bulk reactions and those coupled to a compartment  
are the two-channel or multichannel systems.
In these models, particles are either conserved by the
dynamics in each lane and interaction is realized by the dependence
of the hop rates on the configuration of the parallel lanes
\cite{popkov,peschel,lee,salerno,popkov1,narrow},  
or particles can jump between lanes \cite{pk,harris,mitsudo,reichenbach}. 
We study in this work a two-lane exclusion process where particles
move in the two lanes in opposite directions. 
Particles are allowed to change lanes and we restrict ourselves to the
case of weak lane change rates, i.e. they are inversely proportional
to the system size. This means that the probability that a
marked particle changes lanes during the time it resides in the system
is $\mathcal{O}(1)$. 
If particles in one of the lanes are regarded as holes, and vice versa, 
this model can also be interpreted as a two-channel driven system where
particles move in the same direction in the channels and are
created and annihilated in pairs. 
In the hydrodynamic limit of the model, we shall construct the
steady-state phase diagram  by means of analyzing the differential
equations describing the model on the macroscopic scale.  
At the coexistence line, where coherently moving delocalized shocks
develop in both lanes, which is reminiscent of the delocalized
shock dynamics at the coexistence line of the TASEP, the density
profiles are studied in the framework of a phenomenological domain
wall picture based on the hydrodynamic description.
Recently, a two-lane exclusion process has been investigated with
weak, symmetric lane change, where particles move in the lanes 
in the same direction
\cite{reichenbach}. In this model, the formation of delocalized
shocks in both lanes has been found, as well.
In our model, even the case of asymmetric lane change can be 
treated analytically in the hydrodynamic limit if
the total current is zero, which holds also at the coexistence line. 

The paper is organized as follows. In Sec. \ref{description}, the model
is introduced and the hydrodynamic description is discussed.  
In Sec. \ref{slc}, the case of symmetric lane change is investigated, while 
Sec. \ref{alc} is devoted to the asymmetric case. The results are
discussed in Sec. \ref{discussion} and some of the calculations 
are presented in two Appendixes.   


\section{Description of the model}
\label{description}

\begin{figure}[h]
\includegraphics[width=0.8\linewidth]{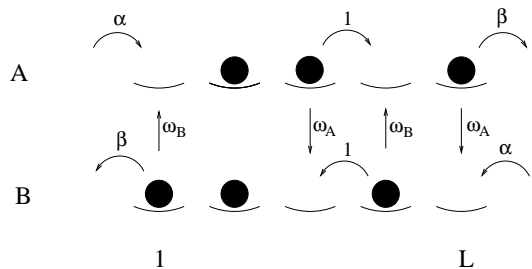}
\caption{\label{model} Transitions and the corresponding
  rates in the model under study.}
\end{figure}

The model we focus on consists of two parallel one-dimensional
lattices with $L$ sites, denoted by $A$ and $B$, the sites of which
 are either empty or occupied
by a particle. 
The state of the system is specified by the set of
occupation numbers $n^{A,B}_i$ which are zero (one) for empty
(occupied) sites. 
We consider in this system a continuous-time stochastic process where
the occupations of pairs of adjacent sites change independently
and randomly after an exponentially distributed waiting time.  
The possible transitions and the
corresponding rates, i.e. the inverses of the mean waiting times,
are the following (Fig. \ref{model}). 
On chain $A$, particles attempt to jump to the adjacent
site on their right-hand side, whereas on chain $B$ to the adjacent
site on their left-hand side, with a rate which is set to unity, and attempts are
successful when the target site is empty.   
On the first site of chain $A$ and on the $L$th site of chain $B$
particles are injected with rate $\alpha$, provided these sites are
empty, whereas on the $L$th site
of chain $A$ and on the first site of chain $B$ they are removed with
rate $\beta$. 
So the system may be regarded to be in contact with
virtual particle reservoirs with densities $\alpha$ and
$1-\beta$ at the entrance- and exit sites, respectively.     
The process described so far is composed of two independent
totally asymmetric simple exclusion processes.
The interaction between them is realized by allowing 
a particle residing at site $i$ of chain $A$($B$) to hop to site $i$ of chain
$B$($A$) with rate $\omega_{A}$ ($\omega_{B}$), provided the target site
is empty. 

As in the case of the TASEP with Langmuir kinetics,
one must distinguish here between three cases, concerning the order
of magnitude of the lane change rates in the large $L$ limit.   
If the rates $\omega_{A}$ and $\omega_{B}$ are of larger order than
$\mathcal{O}(1/L)$, then in the limit $L\to\infty$,  
the interchain processes are dominant compared to the effects of the
boundary reservoirs and the horizontal motion of the particles.
The densities $\rho$ and $\pi$ in lane $A$ and $B$, respectively, 
are expected to be constant far from 
the boundaries and to fulfill the relation  
$\omega_{A}\rho (1-\pi)= \omega_{B}\pi (1-\rho)$, 
which is forced by the lane change kinetics. 
When the interchain hop rates are  smaller than
$\mathcal{O}(1/L)$, then (apart from some possible special parameter combinations\cite{js}) 
they are irrelevant in the $L\to \infty$ limit and the stationary state 
is that of two independent exclusion processes.   
An interesting situation arises   
if the rates $\omega_{A}$ and $\omega_{B}$ are proportional to $1/L$. 
In this case the effects of boundary reservoirs and those of lane
change kinetics are comparable and the
competition between them results in nontrivial density profiles. 
We focus on this case in the present work, and
parametrize the lane change rates as $\omega_{A}=\Omega_{A}/L$ and 
$\omega_{B}=\Omega_{B}/L$, with the constants $\Omega_{A}$ and
$\Omega_{B}$.    
Setting the lattice constant $a$ to $a=1/L$ and 
rescaling the time $t$ as $\tau=t/L$,
we are interested in the properties of the system in the (continuum) limit
$L\to\infty$, where the state of the system is 
characterized by the local densities 
$\rho(x,\tau)$ and $\pi(x,\tau)$ on
chain $A$ and $B$, respectively, which are functions of the continuous
space variable $x\in [0,1]$ and time $\tau$.     
Turning our attention to the subsystem containing lane
$A$($B$) alone, we see that the 
interchain hoppings can be interpreted as 
bulk nonconserving processes for the TASEP in lane $A$($B$). 
The bulk reservoir which the TASEP is connected to is, however, 
not homogeneous but  
it is characterized by the position and time-dependent density $\pi(x,\tau)$($\rho(x,\tau)$).   
Generally, driven diffusive systems which 
are combined with a weak (i.e. $\mathcal{O}(1/L)$)
bulk nonconserving process   
are described on the macroscopic scale specified above by the partial
differential equation 
\be
\frac{\partial \rho (x,\tau)}{\partial \tau}+\frac{\partial J(\rho (x,\tau))}{\partial x}=S(\rho(x,\tau)),
\label{hgen}
\ee  
where  $S(\rho(x,t))$ is the
source term related to the nonconserving process and $J(\rho)$
is the current as a function of the density 
in the steady state of the corresponding translation invariant
infinite system
without nonconserving processes 
(i.e. $S(\rho(x,\tau))\equiv 0$) \cite{schutz03}. 
Under these circumstances, the TASEP has a product measure stationary state
and the current-density relationship is simply $J(\rho)=\rho(1-\rho)$, 
hence 
the currents in the two lanes are 
given as $J_A(\rho(x,\tau))=\rho(x,\tau)[1-\rho(x,\tau)]$ and $J_B(\pi(x,\tau))=-\pi(x,\tau)[1-\pi(x,\tau)]$ in terms 
of the local densities.
For the source terms in the two lanes, we may write
$S_A(\rho(x,\tau),\pi(x,\tau))=-S_B(\rho(x,\tau),\pi(x,\tau))=\Omega_{B}[1-\rho(x,\tau)]\pi(x,\tau)-\Omega_{A}\rho(x,\tau)[1-\pi(x,\tau)]$
since the lane change events at a given site are infinitely rare in the limit
$L\to\infty$.
Setting these expressions into eq. (\ref{hgen}) we obtain that 
in the steady state, where 
$\partial_{\tau}\rho(x,\tau)=\partial_{\tau}\pi(x,\tau)=0$, 
the density profiles $\rho(x)$ and $\pi(x)$
satisfy the coupled differential equations 
\beqn
(2\rho-1)\partial_x\rho +
\Omega_{B}(1-\rho)\pi-\Omega_{A}\rho(1-\pi)=0,  \nonumber \\
(2\pi-1)\partial_x\pi +
\Omega_{B}(1-\rho)\pi-\Omega_{A}\rho(1-\pi)=0.
\label{diff1}
\eeqn 
We mention that one arrives at the same differential equations when in
the master equation of the process the expectation values 
 of pairs of occupation numbers $\langle n_in_{j}\rangle$
are replaced by the products $\langle n_i\rangle \langle n_{j}\rangle$, and
afterwards it is turned to a continuum description with retaining only 
the first derivatives of the densities and neglecting the higher derivatives
which are at most of the order $\mathcal{O}(1/L)$ almost everywhere.  

For the stationary density profiles the boundary conditions 
 $\rho(0)=\pi(1)=\alpha$ and $\rho(1)=\pi(0)=1-\beta$
are imposed. 
In fact, we shall keep these boundary conditions only for
 $\alpha,\beta\le 1/2$; otherwise, we  modify them 
 for practical purposes at the level of the hydrodynamic
 description. The reason for this is the following.
In the domain $\alpha,\beta>1/2$ of the TASEP, the so-called maximum
 current phase, the current is limited by the maximal
  carrying capacity in the bulk, $J=1/4$, which is realized at
 the bulk density $\rho=1/2$ \cite{derrida,schutzdomany}.     
In this phase, boundary layers form in the stationary density profile at both
 ends, where the density drops to the bulk value $\rho=1/2$.
Similarly, in the case of the TASEP with Langmuir kinetics, 
if the entrance rate $\alpha$ exceeds the value $1/2$, 
then in the density profile dictated by the reservoir at the entrance
 site, a boundary layer develops, where the density drops to $1/2$. 
The width of the boundary layer 
is growing sublinearly with $L$ \cite{js}, 
such that in the hydrodynamic limit, it shrinks to $x=0$  
and $\lim_{x\to 0}\rho(x)=1/2$ holds, independently of
 $\alpha$, which influences only the shape of the microscopic
boundary layer. 
These considerations apply also to the present model at both ends and for both
 lanes. 
Therefore, in order to simplify the treatment of the problem
 at the level of the {\it hydrodynamic} description, we use the
 effective boundary conditions   
\be 
\rho(0)=\pi(1)=a, \qquad  \rho(1)=\pi(0)=1-b,
\label{bc}
\ee 
where $a\equiv \min\{\alpha,1/2\}$ and $b\equiv \min\{\beta,1/2\}$.
However, we stress that, although, the profile propagating from e.g. the
left-hand boundary$\rho_l(x)$ is continuous at $x=0$ according to the
effective boundary conditions (\ref{bc}) for $\alpha>0$, a boundary
layer forms on the {\it microscopic} scale.     

In addition to the boundary layers related to the maximal  
carrying capacity in the bulk, the stationary density profiles may
in general contain another type of boundary layer of finite width
or a localized shock in the bulk, where the density has a 
finite variation within a region the width of
which is growing sublinearly with $L$ \cite{frey,ejs}. 
This leads to the appearance of discontinuities in 
$\rho(x)$ and $\pi(x)$ in the hydrodynamic limit, either 
in the bulk $0<x<1$ in the case of a  shock or 
at $x=0,1$ in the case of a boundary layer.
This is in accordance with the fact that, in general, there does not
exist a continuous solution to the two first order differential
equations, which fulfills all four boundary
conditions.   
Apart from some special parameter combinations, there is 
one discontinuity in each lane, which is either in the bulk (a shock)
or at $x=0,1$ (a boundary layer). 
The location of the discontinuity is determined by the requirement
that the currents in both lanes $J_A(\rho(x))$ and $J_B(\rho(x))$ must be 
continuous functions of $x$ in the bulk $0<x<1$ \cite{schutz03,ejs}. 
This follows from that the width of the shock region is proportional 
to $\sqrt{L}$, 
thus the rate of a lane change event is vanishing there in the limit
$L\to\infty$. This condition permits only such a shock 
which separates complementary densities on its two sides, i.e. 
$\rho$ and $1-\rho$ in lane $A$ or $\pi$ and $1-\pi$ in lane $B$.
The position of the shock $x_s$ e.g. in lane A is thus given implicitly
by the equation $\rho_l(x_s)=1-\rho_r(x_s)$, where $\rho_l(x)$ and
$\rho_r(x)$ are the solutions on the two sides of the shock. 
For the detailed rules on the stability of the discontinuity at
$x=0,1$ see Ref. \cite{schutz03}.

Subtracting the two differential equations yields the obvious result that the total
current 
\be
J\equiv \rho(x)[1-\rho(x)]-\pi(x)[1-\pi(x)]
\label{J}
\ee
is a (position independent) constant. 
This relation makes it possible to eliminate one of the
functions, say, $\pi(x)$ and to reduce the problem to the
integration of a single differential equation
\be
\frac{d\rho}{dx}=\Omega_{A}\frac{\rho-[\frac{1}{2} \pm \sqrt{(\rho -\frac{1}{2})^2+J}][K(1-\rho)+\rho]}{2\rho -1},
\label{diffgen}
\ee
where we have introduced the ratio of lane change rates 
$K\equiv \Omega_B/\Omega_A$
and the signs in front of the square root are related to the
two solutions $\pi_+(x)>1/2$ and $\pi_-(x)<1/2$ of the quadratic
equation (\ref{J}). 
Disregarding the simple case $K=1$,
there are two difficulties about this equation. 
First, the solution depends on the current $J$ as a parameter, which itself
depends on the density profiles and is {\it a priori} not known.
Fortunately, apart from two phases in the phase diagram, $\rho(x)$ and 
$\pi(x)$ simultaneously fit to the boundary conditions either at 
$x=0$ or $x=1$, consequently, the current is exclusively determined 
by the entrance- and exit rates as $J=a(1-a)-b(1-b)$. 
In the remaining two phases, the functions $\rho(x)$ and $\pi(x)$
 meet the boundary
conditions at the opposite ends of the system. Here, one may solve
eq. (\ref{diffgen}) iteratively until self-consistency is attained.  
Second, even in the case when $J$ is known, eq. (\ref{diffgen}) cannot be
analytically integrated in general, except for the case when the current 
is zero.  
This is realized in three cases, two of which are related to the
symmetries of the system. We discuss these possibilities in the rest
of the section. 

As the two entrance- and exit rates were chosen to be identical, 
the obvious relation holds when the rates $\Omega_{A}$ and
$\Omega_{B}$ are interchanged: 
\be 
\rho(x;\alpha,\beta,\Omega_{A},\Omega_{B})=
\pi(1-x;\alpha,\beta,\Omega_{B},\Omega_{A}),
\label{sym1}
\ee
where the dependence of the profiles on the four parameters 
$\alpha$,$\beta$,$\Omega_{A}$ and $\Omega_{B}$ is explicitly indicated.
This relation, together with eq. (\ref{J}) implies 
that the current changes sign if $\Omega_{A}$ and $\Omega_{B}$ are
interchanged.  
Thus $\Omega_{A}=\Omega_{B}$ implies $J=0$, that holds apparently  
since none of the chains is singled out in this case. 

As a consequence of the particle-hole symmetry of the model, 
we have the relation
\be
\rho(x;\alpha,\beta,\Omega_{A},\Omega_{B})=
1-\pi(x;\beta,\alpha,\Omega_{A},\Omega_{B}).
\label{sym2}
\ee 
Using eq. (\ref{J}), it follows that the current
changes sign when $\alpha$ and $\beta$ are interchanged, 
so it must be zero for $\alpha=\beta$. 
Alternatively, this can be seen by interchanging particles and holes 
in one of the chains, which results in a two-channel system where 
particles move in the channels in the same direction 
and particles are created
and annihilated in pairs at neighboring sites of the two chains with
rates $\Omega_{A}$ and $\Omega_{B}$, respectively. Particles are injected and 
removed in the channels with the same rate, hence the channels are 
equivalent. Since the currents of particles and holes are equal,
the total current must be zero.

The third parameter regime where the current is zero is the domain 
$\alpha,\beta \ge1/2$.
Here, as aforesaid, the density profiles and the current are 
independent of $\alpha$ and $\beta$ in the hydrodynamic limit. 
Since the current is zero for $\alpha=\beta$ it follows that 
$J=0$ in the whole domain $\alpha,\beta \ge 1/2$.

\section{Symmetric lane change}
\label{slc}

We start the investigation of the model with the simple case 
of equal lane change rates ($\Omega_{A}=\Omega_{B}\equiv\Omega$), 
where the solutions to the hydrodynamic
equations are analytically found and some general features of the model
can be understood. 
Since the total current is zero, either $\rho(x)=\pi(x)$  
or  $\rho(x)=1-\pi(x)$ must hold. 
Substituting the former relation into eq. (\ref{diff1}) yields
\be 
\rho_e(x)=\pi_e(x)={\rm const},
\label{ek1}
\ee
whereas the latter gives 
\be 
\rho_c(x)=1-\pi_c(x)=\Omega x+{\rm const}.
\label{ck1}
\ee
Thus, the profiles
are piecewise linear and consist of constant segments with equal
densities in the two lanes and segments of slope $\Omega$
($-\Omega$) in lane $A$($B$) with complementary densities.    
Switching off the interchain particle exchange ($\Omega=0$), 
we get two identical TASEPs, 
which have, apart from the coexistence line
$\alpha=\beta <1/2$,  constant density profiles in the bulk.
In the high-density phase ($\beta<\min\{\alpha,1/2\}$),
the density, being $1-\beta$, is controlled by the exit rate
 and the profile is discontinuous at $x=0$.
In the low-density phase ($\alpha<\min\{\beta,1/2\}$),
the density is
$\alpha$ and a discontinuity appears at $x=1$ \cite{derrida,schutzdomany}. 
In the maximum current phase ($\alpha,\beta> 1/2$), as we have already
mentioned, the bulk density is $1/2$ and boundary layers appear at
both ends.  
On the other hand, the effect of symmetric lane change processes 
is to diminish the difference between the local densities in the two lanes.  
Since the densities are already equal without the interaction, 
this situation is obviously not altered when switching on the vertical
hopping processes. 
Consequently, the density profiles in the bulk are identical to that
of the TASEP in these phases.
 
\begin{figure}[h]
\includegraphics[width=1.0\linewidth]{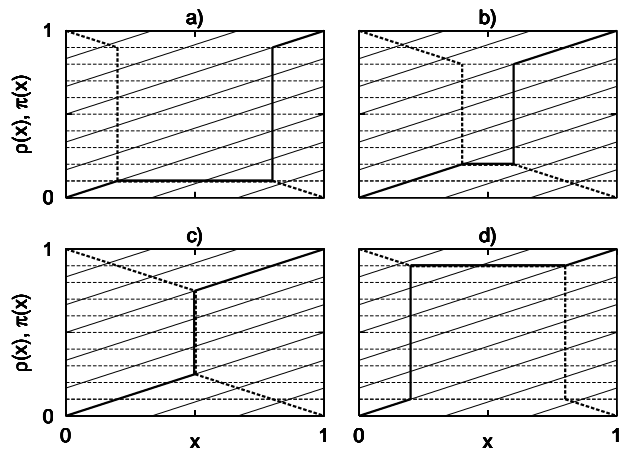}
\caption{\label{k1fig1} Density profiles of the closed system 
 ($\alpha=\beta=0$) with $\Omega=0.5$ for different 
global densities: (a) $\varrho=0.26$, (b) $\varrho=0.44$, (c)
 $\varrho=0.5$ and (d) $\varrho=0.74$. The thin solid and dashed
 lines represent the flow field of the differential equation
 (\ref{diffgen}) corresponding to the complementary-density and
 equal-density solutions, respectively. The thick solid and dashed lines are
 the density profiles $\rho(x)$ and $\pi(x)$, respectively.}
\end{figure}

This is, however not the case at the coexistence line $\alpha=\beta
<1/2$. In the TASEP, a sharp domain wall emerges here in the density
profile, which separates a low- and a high-density phase with 
constant densities far from the domain wall $\alpha$ and $1-\alpha$, 
respectively.
The stochastic motion of the domain wall is
described by an unbiased random walk with reflective boundaries
\cite{schutzdomany,schutz98}, such that the average stationary 
density profile connects linearly the boundary
densities $\alpha$ and $1-\alpha$.   
Returning to our model, we consider first the closed system, 
i.e. $\alpha=\beta=0$. The
profiles which fulfill the requirement about the continuity of the
currents are depicted in Fig. \ref{k1fig1} for various global
particle densities $\varrho\equiv\lim_{L\to\infty}\frac{N}{2L}$, where
$N$ is the number of particles in the system. 
Here, the density profiles consist of three segments in general. In
the middle part of the system an equal-density segment is found
(Fig. \ref{k1fig1}a,b,d). This region is connected with the boundaries
by complementary-density segments on its left-hand side and on its
right-hand side, which    
are continuous at $x=0$ and $x=1$, respectively. 
In both lanes, the density profile is continuous 
at one end of the equal-density segment and a shock is located
at the other one, such that the two shocks are at opposite ends. 
The density in the equal-density region (and at the same time the
location of the shocks) depend on the global particle density.   
At $\varrho =1/2$, the equal-density segment is lacking if $\Omega<1$
(Fig. \ref{k1fig1}c) and the profiles are linear if $\Omega =1$. 

If particles are allowed to enter and exit from the system at the
boundaries, i.e. $0<\alpha=\beta<1/2$, the total number of particles
is no longer conserved. 
Nevertheless, we expect that 
the stationary density profiles averaged over
configurations with a fixed global density $\varrho$, $\rho_{\varrho}(x)$ and $\pi_{\varrho}(x)$,
can still be constructed from the solutions (\ref{ek1}) and (\ref{ck1}) 
of the hydrodynamic equations. These profiles 
are similar to those of the closed system and the only
difference is that the complementary-density segments fit to the
altered boundary conditions 
$\rho_{\varrho}(0)=\pi_{\varrho}(1)=\alpha$ and $\rho_{\varrho}(1)=\pi_{\varrho}(0)=1-\alpha$. 
This is indeed the case in the limit $\alpha\to 0$. Here, 
the injections and removals of particles at the boundary sites, which
modify the global density, are infinitely rare, such that 
the system has always enough time to relax, i.e. to adjust the density
profiles to the slightly altered global density. 
As long as the shocks are not in the vicinity of the boundaries, 
the densities at the boundary sites are independent from the
global density, which influences only the 
position of the equal density segment. 
Therefore, the stochastic variation of the global density $\varrho
(t)$ is described by a homogeneous, symmetric random walk in the 
interval $[0,1]$ with reflective boundary conditions.   
For finite $\alpha$, we can give only a heuristic argument why we
expect that the fluctuations of the global density are quasistationary
in the above sense. 
In the stationary state, the center of the mass of a small instantaneous local
perturbation propagates with a 
velocity $v(x)=1-2\rho(x)$ \cite{lighthill,schutz98}, which changes sign at 
$\rho=1/2$. 
In the complementary-density segments, the perturbations in the
density, which come from the fluctuations of the boundary reservoirs, 
are thus driven toward the equal-density segment with a finite velocity.
The characteristic traveling time of the perturbation, as well as the 
time scale related to the lane change processes in a finite system of size $L$
is $\mathcal{O}(L)$. The relaxation time 
of the perturbation is 
thus expected to be $\mathcal{O}(L)$. However, the random walk
dynamics of the global density implies that the time scale of a finite 
change in the global density is $\mathcal{O}(L^2)$, which is large
compared to the relaxation time, thus the density profile 
has enough time to follow the instantaneous global density.  
The fluctuating global density $\varrho(t)$ is thus expected
to be a symmetric random walk with reflective boundaries at $\alpha$ and
$1-\alpha$.
In the stationary state, the global density is therefore homogeneously
distributed in the interval $[\alpha,1-\alpha]$. 
 
On the other hand, one can easily calculate that 
if the position of the shock in lane $A$ is $x_s$,
the global density of particles in the system is  
$\varrho(x_s)=(1-x_s)[\Delta(x_s)+\Omega]+\alpha$ for $x_s\ge 1/2$, where 
we have introduced the (position dependent) height of the shock: 
$\Delta(x_s)=2\Omega(x_s-1)+1-2\alpha$. Note that, 
as opposed to the single lane
TASEP, this relation is no longer linear, therefore the 
probability distribution of the position of the shock
is not uniform in the steady state. 

\begin{figure}[h]
\includegraphics[width=0.9\linewidth]{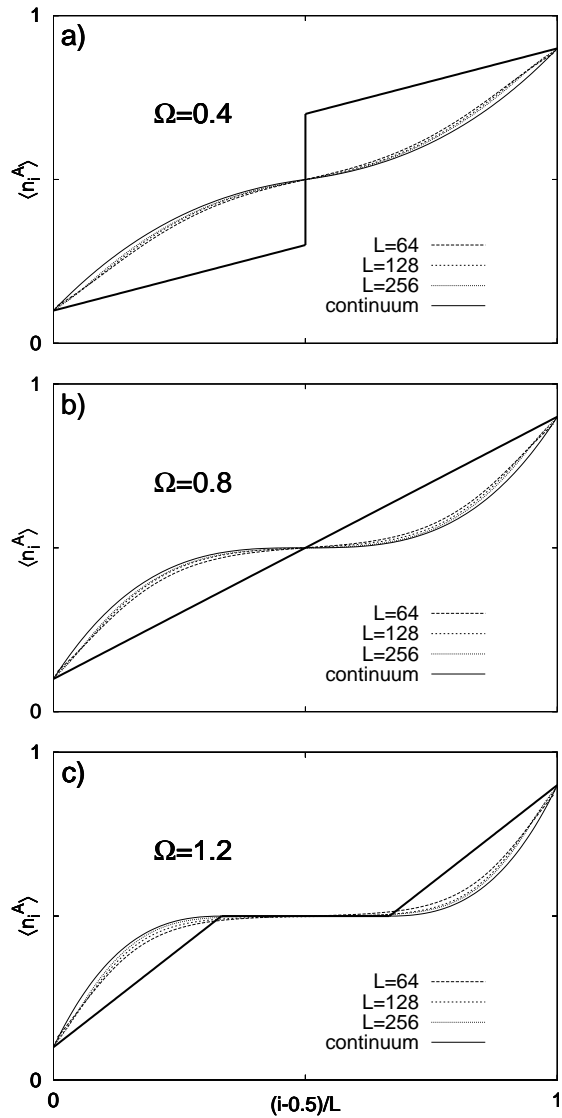}
\caption{\label{k1num4} 
Density profiles in lane $A$ at the coexistence
line at $\alpha=\beta=0.1$, obtained by 
numerical simulation for different system sizes
and for different values of $\Omega$: 
a) $\Omega=0.4$, b) $\Omega=0.8$, and c) $\Omega=1.2$. 
In the case $\Omega=0.8$, the height of the shock at $x_s=1/2$ is zero. 
The solid curves are the analytical predictions in the hydrodynamic limit. 
The thick solid lines represent the profile
$\rho_{\varrho}(x)$ at global density $\varrho=1/2$.}
\end{figure}
With these prerequisites, 
the steady-state density profile in lane A can be easily calculated 
by averaging $\rho_{\varrho}(x)$ over the
steady-state distribution of the global density:  
$\rho(x)=\frac{1}{1-2\alpha}\int_{\alpha}^{1-\alpha}\rho_{\varrho}(x)d\varrho$.
Skipping the details of the straightforward calculations, 
we shall give the profile $\rho(x)$ in the interval 
$\frac{1}{2}\le x \le 1$, whereas for $0 \le x\le \frac{1}{2}$, it
is obtained by the help of the relation
$\rho(x)=1-\rho(1-x)$, that follows from eq. (\ref{sym1}) and
(\ref{sym2}). The density profile in lane $B$ can be calculated by
making use of eq. (\ref{sym2}), which implies $\pi(x)=1-\rho(x)$. 
The cases corresponding to the different signs of $\Delta(\frac{1}{2})$ must
be treated separately. 
For $\Delta(\frac{1}{2})\ge 0$, we obtain
\be 
\rho(x)=\frac{1}{2}+\frac{(x-\frac{1}{2})\Delta^2(x)}{1-2\alpha}  \qquad
\Delta(\frac{1}{2}) \ge 0,
\label{positive}
\ee
which is a third-degree polynomial of $x$. 
If $\Delta(\frac{1}{2})>0$, the second
derivative of $\rho(x)$ is discontinuous at $x=\frac{1}{2}$.
In the limit $\Omega\to 0$, the linear profile of the TASEP at the
coexistence line is recovered.  
If $\Delta(\frac{1}{2})=0$, eq. (\ref{positive}) simplifies to 
\be
\rho(x)=\frac{1}{2}+4\Omega (x-\frac{1}{2})^3 \qquad \Delta(\frac{1}{2})=0,
\ee
which is everywhere analytic. 
For $\Delta(\frac{1}{2})<0$, the profile is constant in the
interval $\frac{1}{2} \le x
\le|\Delta(\frac{1}{2})|/(2\Omega)+\frac{1}{2}$, 
where $\rho(x)=1/2$, while it is given by
eq. (\ref{positive}) in the interval  
$|\Delta(\frac{1}{2})|/(2\Omega)+\frac{1}{2}\le x\le 1$.
These curves, as well as results of Monte Carlo simulations for finite
systems of size $L=64,128$ and $256$ are shown in Fig \ref{k1num4}. 
In the numerical simulations, after waiting a period of $10^6$ Monte Carlo
steps in order to reach the steady state, we have measured the local
occupancies every $10$ Monte Carlo steps during a period of
$5\cdot 10^9$ steps.  
For increasing $L$, the properly scaled profiles seem 
to tend to the analytical curves expected to be valid 
in the continuum limit.

\section{Asymmetric lane change}
\label{alc}

\begin{figure}[h]
\includegraphics[width=1\linewidth]{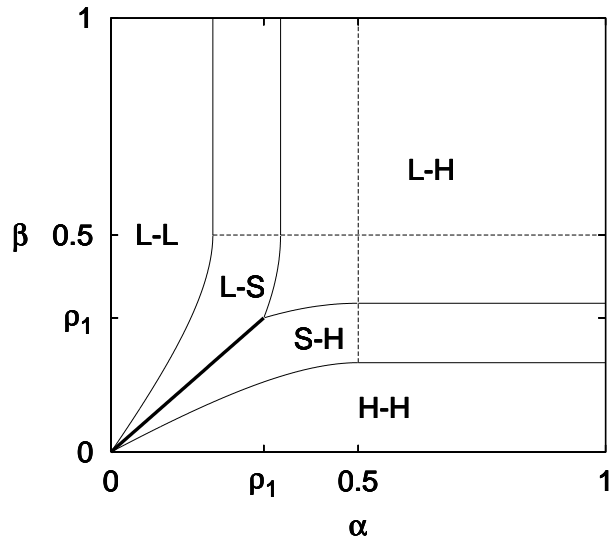}
\caption{\label{pd5} Phase diagram at $\Omega_A=2$,
  $\Omega_B=0.4$. Phase boundaries are indicated by solid lines.
Letters L,H and S refer to low-density,
  high-density and localized shock phase, respectively; the first(second) 
  letter refers to lane $A$($B$).
The thick solid line indicates the coexistence line.  
 At the dashed lines, the function $J(\alpha,\beta)$ is nonanalytic.}
\end{figure}

\begin{figure}[h]
\includegraphics[width=1\linewidth]{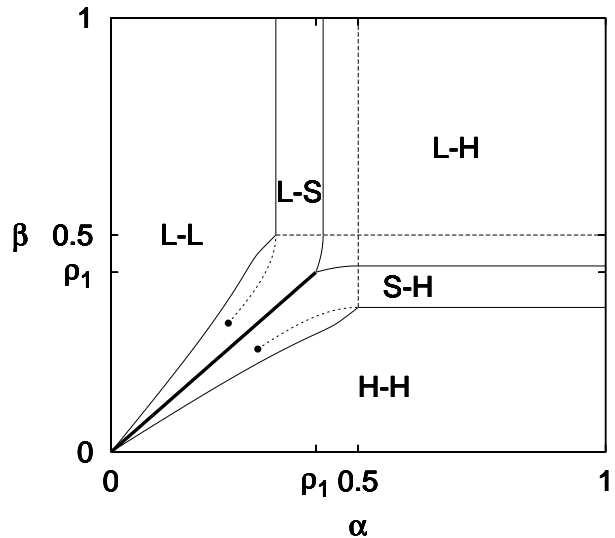}
\caption{\label{pd2} Phase diagram at $\Omega_A=2$,
  $\Omega_B=1$. The dotted curves are the discontinuity lines, which
  terminate at the points indicated by the full circles.}
\end{figure}

In this section, the stationary properties of the model are
investigated in the case $K\neq 1$.
Due to the symmetries of the system, we may restrict ourselves to the 
investigation of the part $K<1$, 
$\beta\le \alpha$ of the parameter space, which is related to the remaining part through eqs. (\ref{sym1}) and (\ref{sym2}).  
Analyzing the solutions to the hydrodynamic equation (\ref{diffgen}), 
one can construct the phase diagram in the four-dimensional parameter
space spanned by $\alpha,\beta,\Omega_A$ and $\Omega_B$. 
Two representative two-dimensional cross sections 
of the parameter space
at fixed lane change rates are shown in
Fig. \ref{pd5} and \ref{pd2}. As can be seen, the phase boundaries are
symmetric to the diagonal, which is a consequence of eq. (\ref{sym2}). 
On the other hand, the phase diagrams are richer compared 
to that of the symmetric model: 
Besides the phases where the profiles are continuous in the interior
of the system, the asymmetry in the lane change kinetics leads 
to the appearance of phases where one of the lanes contains a
localized shock in the bulk. This is reminiscent of the
shock phase of the single lane TASEP with Langmuir
kinetics. As a new feature, the position of the shock may vary
discontinuously with the boundary rates here, when the so-called
discontinuity line is crossed. The coexistence line, where 
coherently moving delocalized shocks emerge in both lanes, is still present,
however, it is shorter than in the symmetric case and the shocks
walk only a shrunken domain. 
The subsequent part of the section is devoted to the detailed analysis of
these findings.   

\subsection{Density profiles}

We start the presentation of the results with the description of the
density profiles in the phases below the diagonal $\alpha=\beta$ of
the two-dimensional phase diagrams.
 
\begin{figure}[h]
\includegraphics[width=1\linewidth]{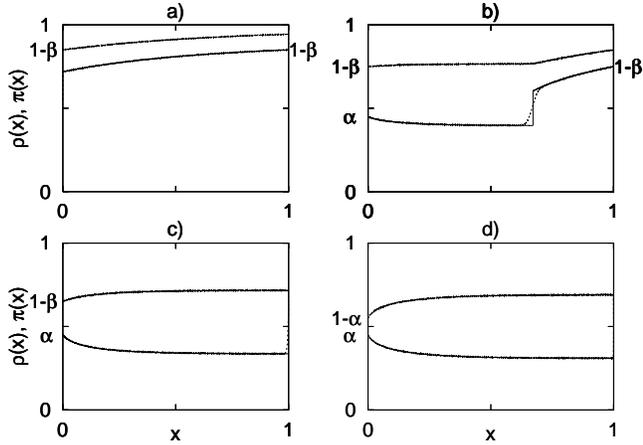}
\caption{\label{4prof} Density profiles in lane $A$ (lower line) and lane
 $B$ (upper line) for the parameters $\Omega_A=2$, $\Omega_B=0.4$,
 $\alpha=0.45$ and for various exit rates: a) $\beta=0.15$, b) $\beta=0.25$,
 c) $\beta=0.35$ and d) $\beta=0.45$. The profiles in panels a)-c)
 were obtained by numerically solving eq. (\ref{diffgen}), whereas those
 in panel d) are the analytical curves in eq. (\ref{comp}). Results of
 numerical simulations (dotted lines) obtained for system size
 $L=10000$ by averaging the occupancies in a period of $10^7$ Monte Carlo steps
 in the steady state are
 hardly distinguishable from these curves.}
\end{figure}

If the exit rate is small enough, the densities in the bulk 
exceed the value $1/2$ in both lanes (Fig. \ref{4prof}a). 
Both profiles are continuous in the bulk and at the
exits, i.e. $\lim_{x\to 1}\rho(x)=\lim_{x\to 0}\pi(x)=1-b$,  
but they are discontinuous at the
entrances, i.e. $\lim_{x\to 0}\rho(x)\neq a$ and $\lim_{x\to
  1}\pi(x)\neq a$, which signals the appearance of boundary
layers on the microscopic scale. 
The profiles $\rho(x)$ and $\pi(x)$, as well as the current, depend 
exclusively on $\beta$, while $\alpha$
influences only the boundary layers at the entrances. This
situation is observed also in the high-density phase of the TASEP with
Langmuir kinetics \cite{frey,ejs}, 
therefore we call this phase H-H phase, referring to 
the high density in both lanes. 

In the phase denoted by S-H in the phase diagram,
$\rho(x)$ and $\pi(x)$ are continuous at the exits, similarly to the
H-H phase, however, the discontinuity in $\rho(x)$ is no longer
at $x=0$ but it is shifted to the interior of the system (Fig. \ref{4prof}b). 
Thus $\lim_{x\to 0}\rho(x)=a$ holds and a shock is located in
lane $A$ at some $x_s$ ($0<x_s<1)$. 
Therefore this phase is termed S-H phase, where the letter S refers to the shock 
in lane $A$ and letter H refers to the high density ($\pi(x)>1/2$)     
in lane $B$. The function $\pi(x)$
is discontinuous at $x=1$, i.e. $\lim_{x\to 1}\pi(x)\neq a$ 
and it is not differentiable (although continuous) at $x_s$.
Since both $\rho(x)$ and $\pi(x)$ are continuous at $x=0$, the 
total current is given by 
\be 
J=a(1-a)-b(1-b)
\label{Jexact}
\ee   
in this phase.
The profile $\rho(x)$ at fixed $\alpha$ and $\beta$  
can be computed by substituting the current calculated from
eq. (\ref{Jexact}) into the differential equation (\ref{diffgen}). 
Then, the solutions propagating from the left-hand and the right-hand
boundary, i.e. the solutions $\rho_l(x)$ and $\rho_r(x)$  
fulfilling the boundary conditions $\rho_l(0)=a$ and
$\rho_r(1)=1-b$, respectively, are calculated numerically. 
Finally, the position of the shock $x_s$ is obtained from 
the relation $\rho_l(x_s)=1-\rho_r(x_s)$, which is 
implied by the continuity of the current in lane $A$.
Once $\rho(x)$ is at our disposal, $\pi(x)$ can be calculated from eq. (\ref{J}).

Apart from the discontinuity line to be discussed in the next section, 
the position of the shock $x_s$ varies continuously with the boundary rates
in the S-H phase.  
Fixing $\alpha$ and reducing $\beta$, $x_s$ is
decreasing and at a certain value of $\beta$, $\beta=\beta_H(\alpha)$, 
the shock reaches the left-hand boundary at $x=0$. 
At this point, the right-hand solution $\rho_r(x)$ extends entirely to
the left-hand boundary and a further increase in $\beta$ drives the system to 
the H-H phase. 
The phase boundary $\beta_H(\alpha)$ between the S-H and the H-H
phase is thus determined from the condition $x_s=0$ or, equivalently, 
$\rho_r(1)=1-a$.  
When $\beta$ is increased along a vertical path in the phase
diagram at a fixed $\alpha$, $x_s$ increases and  
for $\alpha<\rho_1$, where the constant $\rho_1$ will be determined
later, the path hits the coexistence line before the shock would reach
the right-hand boundary (see Sec. III. D).
Increasing $\beta$ along a path at some $\alpha>\rho_1$, 
the shock reaches the right-hand boundary at $x=1$ for a certain value 
of $\beta$, $\beta =\beta_L(\alpha)$ and the path leaves the S-H phase.  
At the phase boundary, 
the left-hand solution $\rho_l(x)$ extends to the whole system 
and $\rho_l(1)=b$ must hold. 

Crossing the phase boundary $\beta_L(\alpha)$, the L-H phase is
entered, where letter L refers to the low density in lane $A$ since
here, $\rho(x)<1/2$ and $\pi(x)>1/2$ hold in the bulk (Fig. \ref{4prof}c).   
In this phase, $\rho(x)$ and $\pi(x)$ are discontinuous at $x=1$, whereas
they are continuous at $x=0$ hence the
current is given by eq. (\ref{Jexact}).  
In the part of the L-H phase where the current is zero, i.e. if
$\alpha=\beta$ or $\alpha,\beta\ge 1/2$, the profiles 
can be calculated analytically. 
The equal-density solutions of eq. (\ref{diffgen}) 
are 
\beqn 
\ln [\rho_e(x)(1-\rho_e(x))]&=&\Omega_{A}(K-1)x+{\rm const}, 
\nonumber \\
\pi_e(x)&=&\rho_e(x),
\label{equal}
\eeqn  
whereas the complementary-density solutions read as  
\beqn 
\frac{\rho_1}{\rho_2}\ln |\rho_c(x)-\rho_1|-\frac{\rho_2}{\rho_1}\ln
|\rho_c(x)-\rho_2|&=&2\Omega_A\sqrt{K}x+{\rm const}, \nonumber \\  
\pi_c(x)&=&1-\rho_c(x),
\label{comp}
\eeqn
where the constants $\rho_1\equiv 1/(1+K^{-1/2})$ and 
$\rho_2\equiv 1/(1-K^{-1/2})$
are the roots of the equation $S_A(\rho,1-\rho)=0$.
There is, furthermore, a special complementary-density solution with
constant densities:
\beqn 
\rho_c(x)&=&\rho_1, \nonumber \\
  \pi_c(x)&=&1-\rho_1.
\label{compsp}
\eeqn 
In the part of the L-H phase where $J=0$, the profiles are given by
the complementary-density solution which fulfills
the boundary conditions $\rho_c(0)=a$ and $\pi_c(0)=1-a$ (Fig. \ref{4prof}d). 

\subsection{Phase boundaries and the discontinuity line}

In the S-H phase (and the L-H phase), the profiles
$\rho(x)$ and $\pi(x)$, as well as the
current are independent of $\alpha$  if $\alpha\ge 1/2$. Here, $\lim_{x\to 0}\rho(x)=1/2$  
and $\alpha$ influences only the microscopic boundary layer, as we
argued in Sec. II. As a
consequence, the phase boundaries $\beta_H(\alpha)$ and
$\beta_L(\alpha)$ are horizontal lines in the domain 
$\alpha\ge 1/2$ (see Fig. \ref{pd5} and \ref{pd2}) and 
we may restrict the investigation of the phase boundaries
to the domain $\alpha\le 1/2$.  

Although we cannot give an analytical expression for the density profiles in
general, some
information can be gained on the phase boundaries of the S-H phase 
by investigating the constant solutions of the hydrodynamic equations.
A constant solution $\rho(x)=r$, $\pi(x)=p$ must obey 
$S_A(r,p)=0$, otherwise the spatial derivatives 
$\partial_x\rho(x)$ and $\partial_x\pi(x)$
would not
vanish in eq. (\ref{diff1}). On the other hand, the constants satisfy
the equation $J=r(1-r)-p(1-p)$, where $J$ is
determined by the boundary rates via eq. (\ref{Jexact}). 
Eliminating $p$ yields that $r$ 
is given by the implicit equation:
\be
r(r-1)\left[1-\frac{K}{(K+(1-K)r)^{2}}\right]=J(\alpha,\beta).
\label{Jrho}
\ee   
In the S-H phase, this equation has two roots $r_1(J(\alpha,\beta),K)$ and 
$r_2(J(\alpha,\beta),K)$ (shortly $r_1$ and $r_2$) 
in the interval $[0,1]$. One can check that for the
larger root $r_2$, the relation $1/2<1-\beta<r_2$ holds, whereas
the smaller one, $r_1$, may be larger or smaller than $1/2$.    

First, we examine the phase boundary separating the L-H phase and the
S-H phase. One can check that at this boundary line, $r_1<1/2$
holds.  Moreover, it follows from eq. (\ref{diff1}) that
$\frac{d\rho(x)}{dx}> 0$ if $0\le\rho(x)<r_1$, and 
$\frac{d\rho(x)}{dx}< 0$ if $r_1<\rho(x)<1/2$. 
Thus, the line $\rho(x)=r_1$ behaves as an attractor for the 
solutions $\rho_l(x)$ propagating from the left-hand boundary  
$x=0$ if $\rho_l(0)=\alpha\le 1/2$, 
meaning that $\rho_l(x)$ approaches monotonously to $r_1$ as $x$ increases and $\lim_{x\to\infty}\rho_l(x)=r_1$. 
Since $\rho_l(x)$ is monotonous and $\rho_l(0)=\alpha$, as well as $\rho_l(1)=\beta$ hold at the phase boundary, at the common point of
the boundary line and the diagonal $\alpha=\beta$, $\rho_l(x)$ must
be a constant function $\rho_l(x)=\alpha$. This, however, implies that $\alpha$
must coincide with $r_1$. The endpoint of the boundary line
$\beta_L(\alpha)$ is therefore at
$\alpha=\beta=r_1(J=0,K)=1/(1+K^{-1/2})$, which depends only on $K$.  

\begin{figure}[h]
\includegraphics[width=0.9\linewidth]{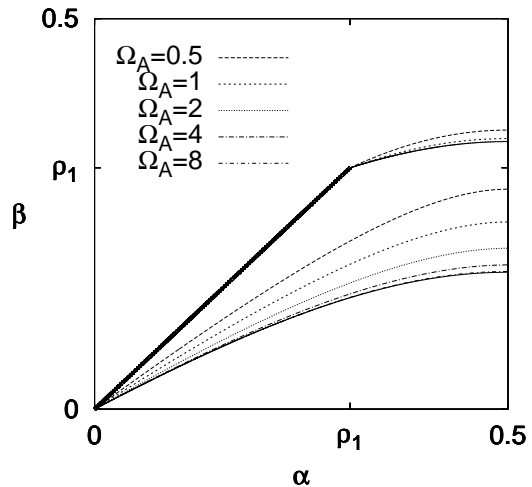}
\caption{\label{phase5} Phase boundaries of the S-H
  phase obtained by integrating eq. (\ref{diffgen}) numerically for $K=1/5$ and for various $\Omega_A$. 
The solid lines are the limiting
  curves: $\beta_{L\infty}(\alpha)$, given by eq. (\ref{pb2}) and 
  $\beta_{H\infty}(\alpha)$, given solely by eq. (\ref{pb1}) since $K<K^*$.}
\end{figure}

As opposed to this point, the whole function $\beta_L(\alpha)$ depends both on 
$\Omega_A$ and $\Omega_B$.  
Nevertheless, we can find an analytical expression for $\beta_L(\alpha)$
 in the limit $K={\rm const}$, $\Omega_A\to\infty$. 
We can see from eq. (\ref{diffgen}) that the
derivative $\frac{d\rho(x)}{dx}$ is proportional to $\Omega_A$ for
a fixed $K$. As a consequence, the larger $\Omega_A$ is the more rapidly 
$\rho_l(x)$ tends to $r_1$. Thus, in the limit specified above, 
$\lim_{\Omega_A\to\infty}(\rho_l(1)-r_1)=0$, which leads to $\beta=r_1$. 
Substituting this into eq. (\ref{Jrho}), we
obtain for the inverse of the boundary curve $\beta_{L\infty}(\alpha)$
in the limit $K={\rm const}$, $\Omega_A\to\infty$:
\beqn 
[\beta_{L\infty}]^{-1}(\beta)= \qquad \qquad \qquad \qquad \qquad
\qquad \qquad  \qquad& & \nonumber \\
\frac{1}{2}\left(1-\sqrt{1-4\beta(1-\beta)
\left[2-\frac{K}{\left[K+(1-K)\beta\right]^2}\right]}\right).& &
\label{pb2}
\eeqn  
This curve is plotted in Fig. \ref{phase5}. 
The phase boundaries obtained by integrating eq. (\ref{diffgen}) 
numerically for 
finite lane change rates tend rapidly to this limiting curve. 

Next, we turn to examine the boundary curve between the S-H and
the H-H phase, $\beta_H(\alpha)$. Along this line,
$\rho_r(0)=1-\alpha$ and $\rho_r(1)=1-\beta$ hold, 
and the roots of eq. (\ref{Jrho}) are
arranged as $r_1<1-\alpha<1-\beta<r_2$. 
One can show that 
the line $\rho(x)=r_2$ is an attractor for the 
solutions $\rho_l(x)$ which start at $x=0$ if 
$\rho_l(0)> \max\{1/2,r_1\}$. Moreover, if $r_1>1/2$, the line
$\rho(x)=r_1$ repels the solutions $\rho_l(x)$ starting from $x=0$ if
$r_1<\rho_l(0)<r_2$, or, in other words, the solutions $\rho_r(x)$ 
propagating from the right-hand boundary, for which $r_1<\rho_r(1)<r_2$, 
are attracted by the line $\rho(x)=r_1$.
When the diagonal is approached along the phase
boundary $\beta_L(\alpha)$, the profile $\rho_r(x)$, being monotonous,
must tend to the constant function $\rho_r(x)=1-\alpha$, as well as
$\pi(x)$ since the current is zero at the diagonal. 
However, equal densities in the two lanes are possible for $K\neq 1$
only if the density is one (or zero), 
therefore the boundary line must approach the
diagonal at $\alpha=0$. This is in accordance with the fact that
$r_2=1$ if $J=0$.
Thus, we obtain $\lim_{\alpha\to 0}\beta_L(\alpha)=0$, independently
of the lane change rates.

\begin{figure}[h]
\includegraphics[width=0.9\linewidth]{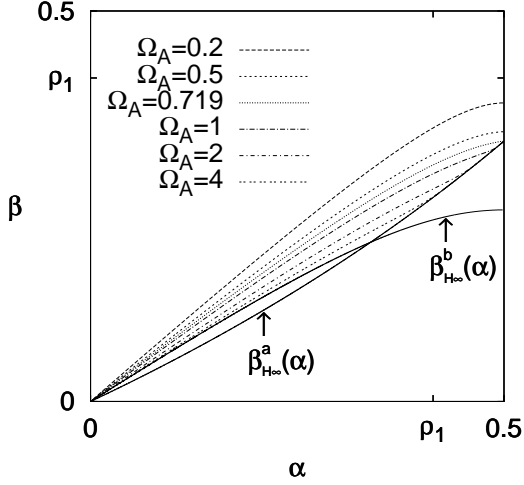}
\caption{\label{phase2} Phase boundaries
  between the S-H phase and the H-H phase obtained by numerical
  integration of eq. (\ref{diffgen})
  for $K=1/2>K^*$  and for several values of $\Omega_A$. The solid lines are
  the limiting curves given in eq. (\ref{pb1}) and eq. (\ref{pb3}).}
\end{figure}

Similarly to $\beta_L(\alpha)$, the location of 
the whole boundary line $\beta_H(\alpha)$
depends on both lane change rates (see Fig. \ref{phase5} and \ref{phase2})
and we can give an analytical expression for $\beta_H(\alpha)$ 
again only in the limit $K={\rm const}$, $\Omega_A\to\infty$. 
 As aforesaid, the profile given by $\rho_r(x)$ lies between two
attractors, $\rho(x)=r_1$ and $\rho(x)=r_2$, to which 
the solutions tend in the limit $x\to-\infty$ and $x\to\infty$, respectively.
When $\Omega_A$ is increased (such that $K$ is fixed), then
$\beta_L(\alpha)$ at a fixed $\alpha$ decreases. 
Thus, the current is increasing and the two roots of
eq. (\ref{Jrho}), $r_1$ and $r_2$ are coming closer.  
Keeping in mind that $r_1<1-\alpha<1-\beta<r_2$, 
there are now two possible cases. Depending on the value of $\alpha$, 
either the gap between $1-\beta$ and $r_2$ or the gap between 
$r_1$ and $1-\alpha$ vanishes first.
In other words, in the former case, the profile is attracted to the
line $\rho(x)=r_2$ at $x=1$ in the limit $\Omega_A\to\infty$, 
i.e.  $\lim_{\Omega_A\to\infty}(\rho_r(1)-r_2)=0$, whereas in the
latter case it is attracted to the line $\rho(x)=r_1$ at $x=0$
(provided that $r_1>1/2$) , i.e.
$\lim_{\Omega_A\to\infty}(\rho_r(0)-r_1)=0$. 
In the first case, substituting $r=1-\beta$ into eq. (\ref{Jrho}), 
we obtain for the inverse of the limiting curve $\beta_{H\infty}^b(\alpha)$
in terms of $\beta_{L\infty}(\alpha)$
\be 
[\beta_{H\infty}^b]^{-1}(\beta)=[\beta_{L\infty}]^{-1}(1-\beta), 
\label{pb1}
\ee  
whereas in the second case, $r=1-\alpha$ yields  
\beqn 
\beta_{H\infty}^{a}(\alpha)= \qquad  \qquad \qquad \qquad \qquad \qquad \qquad  \nonumber \\
\frac{1}{2}\left(1-\sqrt{1-\frac{4\alpha(1-\alpha)K}
{\left[K+(1-K)(1-\alpha)\right]^2}}\right).
\label{pb3}
\eeqn  
These curves are plotted in Fig. \ref{phase2}. 
The phase boundary in the limit $K={\rm const}$, $\Omega_A\to\infty$ is 
given by 
\be 
\beta_{H\infty}(\alpha)=\max\{\beta_{H\infty}^a(\alpha),\beta_{H\infty}^b(\alpha)\}.
\label{env}
\ee
The value $\alpha^*$ at which the functions $\beta_H^a(\alpha)$ and
$\beta_H^b(\alpha)$ intersect varies with $K$. If $K\to 1$,
$\alpha^*$ tends to $\frac{\sqrt{3}-1}{2\sqrt{3}}=0.21132\dots$,
while $\alpha^*=1/2$ if $K$ is equal to  
\be
K^*=1+\sqrt{2}-\sqrt{2(1+\sqrt{2})}=0.21684\dots
\label{kstar}
\ee
Thus, for $K\le K^*$, the limiting curve of $\beta_H(\alpha)$ is given
by eq. (\ref{pb1}) alone, 
otherwise it is composed of eq. (\ref{pb1}) and eq. (\ref{pb3}) as
given by eq. (\ref{env}).   

Although the curve $\beta_{H\infty}(\alpha)$ gives the phase boundary line
only in the limit $K={\rm const}$, $\Omega_A\to\infty$, 
we show that for
$K>K^*$ and for large enough $\Omega_A$, $\Omega_A\ge \Omega_A^*(K)$,
the phase transition point at $\alpha=1/2$ is given exactly by   
eq. (\ref{pb3}), i.e.  $\beta_H(1/2)=\beta^a_{H\infty}(1/2)=\frac{K}{1+K}$.  

\begin{figure}[h]
\includegraphics[width=1.0\linewidth]{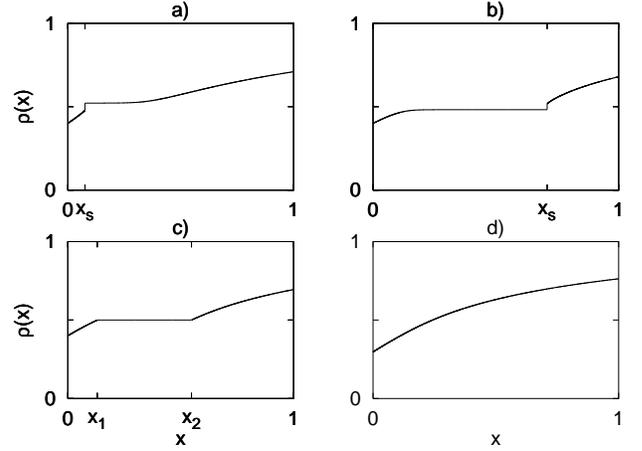}
\caption{\label{crit} Density profiles for parameters $\Omega_A=2$,
  $\Omega_B=1$ and  a) $\alpha=0.4$, $\beta=0.29$, where $r_1$ is slightly
  above $1/2$; b) $\alpha=0.4$, $\beta=0.32$, where $r_1$ is slightly
  below $1/2$; c) $\alpha=0.4$, $\beta\approx 0.305$, where
  $r_1=1/2$. 
  d) The endpoint of the discontinuity line at $\alpha\approx 0.297$, $\beta \approx 0.237$.}
\end{figure}

In order to see this, we discuss first the possible
appearance of a discontinuity line in the S-H phase, at which 
the position of the shock $x_s$ changes discontinuously.
If $r_1=1/2$, one can see from eq. (\ref{diff1}) 
that for the spatial derivative of the profile,
$\lim_{\rho\to 1/2}\frac{d\rho(x)}{dx}>0$ holds and the left-hand and
right-hand solutions $\rho_l(x)$ and $\rho_r(x)$ may propagate as far as the line 
$\rho(x)=1/2$. If $\rho_l(x_1)=1/2$ and $\rho_r(x_2)=1/2$ hold for some
$x_1$ and $x_2$, such that $0<x_1<x_2<1$, then the left-hand and right-hand solutions are connected by  a
constant segment $\rho(x)=1/2$ in the interval $[x_1,x_2]$ and the
profile is continuous (Fig. \ref{crit}c).  
Substituting $r=1/2$ into eq. (\ref{Jrho}) we obtain the equation
of the discontinuity line:  
$\alpha(1-\alpha)-\beta(1-\beta)=(\frac{1-K}{2(1+K)})^2$,
along which the current is constant. Solving this equation for
$\beta$, we obtain 
\beqn
\beta_{d}(\alpha)= \qquad  \qquad \qquad \qquad \qquad \qquad \qquad  \nonumber \\
\frac{1}{2}\left(1-\sqrt{1-4
\alpha(1-\alpha)+4\left(\frac{1-K}{2(1+K)}\right)^2}\right).
\label{disc}
\eeqn
 
This curve is shown in Fig. \ref{pd2}.
When at an arbitrary point of this line, $\beta$ is infinitesimally
decreased, then $r_1$ exceeds the value $1/2$ and a shock
appears at $x_1$ with an infinitesimal height (Fig. \ref{crit}a). 
Conversely, an
infinitesimal increase in $\beta$ decreases $r_1$ below $1/2$ 
and an infinitesimal shock appears at $x_2$ (Fig. \ref{crit}b). 
Thus, when this line is crossed, the position of the shock jumps from
$x_1$ to $x_2$. 
At the point of the discontinuity line at $\alpha=1/2$, $x_1=0$ holds and
an infinitesimal increase (decrease) in $\beta$ drives the system
to the H-H (S-H) phase. Therefore the point of this curve
at $\alpha=1/2$ 
coincides with the phase boundary, i.e. $\beta_d(1/2)=\beta_H(1/2)$. 
On the other hand, comparing eq. (\ref{pb3}) and eq. (\ref{disc}), we
see that $\beta_d(1/2)=\beta_{H\infty}^a(1/2)=\frac{K}{1+K}$. 
The function $\beta_{H\infty}^a(\alpha)$ thus gives the phase transition point 
at $\alpha=1/2$ exactly, provided the curve $\beta_d(\alpha)$ 
is located in the S-H phase. 
That means if $K>K^*$ and $\Omega_A\ge \Omega_A^*(K)$, where 
$\Omega_A^*(K)$ is the value of $\Omega_A$ at which $\beta_H(1/2)$ first 
reaches the value $\frac{K}{1+K}$, when $\Omega_A$ is
increased from zero. For example, for $K=1/2$, we have found $\Omega_A^*(1/2)\approx
0.719$ (Fig. \ref{phase2}). 
We emphasize, however, that the discontinuity line is lacking
if $K<K^*$ or $K>K^*$ but $\Omega_A<\Omega_A^*(K)$.

If $K>K^*$ and $\Omega_A> \Omega_A^*(K)$, then, at the transition point at
 $\alpha=1/2$, we have $x_1=0$; $\Omega_A$ influences only  $x_2$ 
and $x_2\to0$ if $\Omega_A\to\Omega_A^*(K)$.    
Moving away from the point at $\alpha=1/2$ along the discontinuity line, 
the length of the constant segment
$x_2-x_1$ is decreasing and vanishes at a certain point. 
Here, the density profile becomes analytical and the
 discontinuity line terminates (see Fig. \ref{crit}d and Fig. \ref{pd2}). 
The position of this endpoint depends on $\Omega_A$ and it is moving 
toward smaller values of $\alpha$ for increasing $\Omega_A$; 
in the large $\Omega_A$ limit, it tends to the 
point of intersection of $\beta_d(\alpha)$ and $\beta_{H\infty}^b$, whereas 
in the limit $\Omega_A\to\Omega_A^*(K)$, the abscissa of the end point
 tends to $1/2$. 

Finally, we mention that another special curve in the S-H phase 
is defined by the equation $S_A(\alpha,1-\beta)=0$. 
At this curve the left-hand solution $\rho_l(x)$ is constant,
therefore both $\rho(x)$ and $\pi(x)$ are constant in the interval $[0,x_s]$.

\subsection{Current}

We have seen that, apart from the H-H (and the L-L phase), the current
is given by eq. (\ref{Jexact}). It is zero at the line $\alpha=\beta$ and 
 in the domain $\alpha,\beta\ge 1/2$ and it is
 independent of $\alpha$ ($\beta$) in the H-H (L-L) phase. 
It is a continuous function of the boundary rates everywhere, although 
nonanalytic at the boundaries of the H-H phase and the L-L phase, 
as well as at the lines $\alpha=1/2$ and $\beta=1/2$
outside the H-H phase and the L-L phase, 
where the effective boundary rates $a$ and $b$ saturate at $1/2$.   

\begin{figure}[h]
\includegraphics[width=1.0\linewidth]{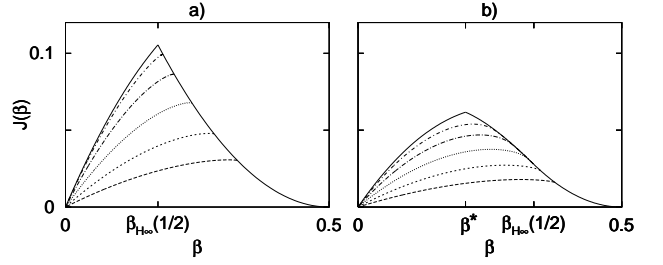}
\caption{\label{current} The current as a function of $\beta$ along
  the line $\alpha=1/2$ for $K=1/5$ (a) and $K=1/2$ (b). The solid
  line is the current in the limit $\Omega_A\to\infty$, the other
  curves from bottom to top correspond to $\Omega_A=0.25,0.5,1,2,4$, respectively.}
\end{figure}

In the following, we concentrate on the current in the H-H phase. 
Since it is continuous at the boundary line
$\beta_H(\alpha)$, it can be expressed in the H-H phase 
in terms of $\beta_H(\alpha)$ as
$J(\beta)=\beta_H^{-1}(\beta)(1-\beta_H^{-1}(\beta))-\beta(1-\beta)$.
According to numerical results (see Fig. \ref{current}), the current $J(\beta)$ has a maximum in
the H-H phase, and for large $\Omega_A$, the location of the maximum tends to $\beta=\beta_H(1/2)$ 
for $K\le K^*$, while for $K>K^*$, it tends to $\beta^*$ where the curves 
$\beta_{H\infty}^a(\alpha)$ and $\beta_{H\infty}^b(\alpha)$ intersect.
Contrary to this, the current is a monotonously decreasing function of $\beta$ in
the S-H phase.
We now turn to the question, at which parameter combination the total current
is maximal in the steady state. 
For fixed $K$, the maximal current is realized in the limit
$\Omega_A\to\infty$ at $\beta_{H\infty}(1/2)$ for $K<K^*$ and at $\beta^*$ 
for $K>K^*$. 
Since the current is growing faster in the S-H phase for decreasing
$\beta$ than in the H-H phase above $\beta^*$ (if $K>K^*$), 
we conclude that the parameter 
combination that maximizes the current is found in the domain $K<K^*$.
For $K<K^*$, the maximal current is thus  
$J_{\rm max}(K)=1/4-\beta_{H\infty}(1/2)(1-\beta_{H\infty}(1/2))$. 
Since $\beta_{H\infty}(1/2)$ decreases monotonously with decreasing $K$, the
current is maximal at $K=0$, where
$\beta_{H\infty}(1/2)=\frac{2-\sqrt{2}}{4}$ and $J_{\rm max}(0)=1/8$. 
The value $1/8$ is thus an upper bound for
the total current and $J\to 1/8$ in the limit $\Omega_A\to \infty$ if
$\alpha\ge 1/2$, $\beta=\frac{2-\sqrt{2}}{4}$ and  $\Omega_B=0$.

\begin{figure}[h]
\includegraphics[width=1.0\linewidth]{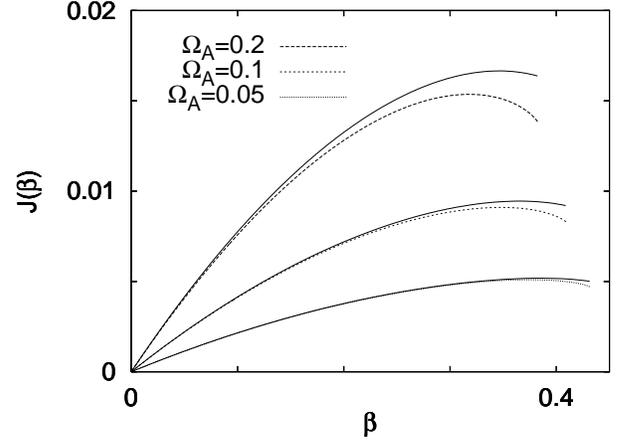}
\caption{\label{approx} The current as a function of $\beta$
  in the H-H phase obtained by numerical integration for $K=1/2$ and for 
different values of $\Omega_A$. The analytical approximation 
given in eq. (\ref{smallJ}) is indicated by solid lines.}
\end{figure}
 
We close this section with the discussion of the current in the H-H
phase when the lane change rates are small, i.e. $\Omega_A,\Omega_B \ll 1$. 
If the vertical hopping processes are switched off, the current is zero, hence 
we expect that for small lane change rates the current is small, as well.  
Assuming that $J\ll (\frac{1}{2}-\beta)^2$ and expanding the right-hand side
of eq. (\ref{diffgen}) in a Taylor series up to first order in $J$, then
solving the resulting differential equation yields finally
\beqn 
J=\frac{1-K}{1+K}\beta(1-\beta)(1-2\beta)\times \nonumber \\
\left[\sqrt{1+2(1+K)\frac{\Omega_A}{1-2\beta}}-1\right]+
\mathcal{O}(\Omega_A^3).
\label{smallJ}
\eeqn
The details of the calculation are presented in Appendix A. 
This expression is compared to the current calculated by integrating
eq. (\ref{diffgen}) numerically in 
Fig. \ref{approx}.
Expanding this expression for small $\Omega_A$, we obtain 
\be 
J=\beta(1-\beta)(1-K)\Omega_A\left[1-\frac{1+K}{2}\frac{\Omega_A}{1-2\beta}\right]
+ \mathcal{O}(\Omega_A^3).
\label{smallJ2}
\ee   
The current is thus in leading order proportional to $\Omega_A$. 
Examining the higher-order terms in the series expansion of the
right-hand side of eq. (\ref{diffgen}), one
can show that for arbitrary $\Omega_A$, the current vanishes 
as $J\sim 1-K$ when $K\to 1$.

\subsection{Coexistence line} 

Let us assume that a given point of the section $\alpha=\beta<\rho_1$
is approached along a path in the S-H phase. 
In this case, the position of the shock in lane $A$, $x_s$, tends to some
$x_{\rm min}=x_{\rm min}(\alpha,\Omega_A,\Omega_B)$, which is somewhere
in the bulk,
i.e. $0<x_{\rm min}<1$. 
When the same point is approached
from the L-S phase, the position of the shock in lane $B$ tends
to the same $x_{\rm min}$ according to eq. (\ref{sym2}).   
Thus, when the boundary line $\alpha=\beta<\rho_1$ is 
passed from the S-H phase, the shock in
lane $A$ jumps from $x_{\rm min}$ to the right-hand boundary at $x=1$, where a discontinuity appears, while 
the discontinuity at $x=1$ in lane $B$, which
can be regarded as a shock localized there, jumps to $x_{\rm min}$.   
So, the density profile changes discontinuously. 
Strictly at $\alpha=\beta$, the shocks in both lanes are delocalized 
and perform a stochastic motion in the domain $[x_{\rm min},1]$, 
similarly to the symmetric model with $K=1$ at the line $\alpha=\beta<1/2$. 

\begin{figure}[h]
\includegraphics[width=1.0\linewidth]{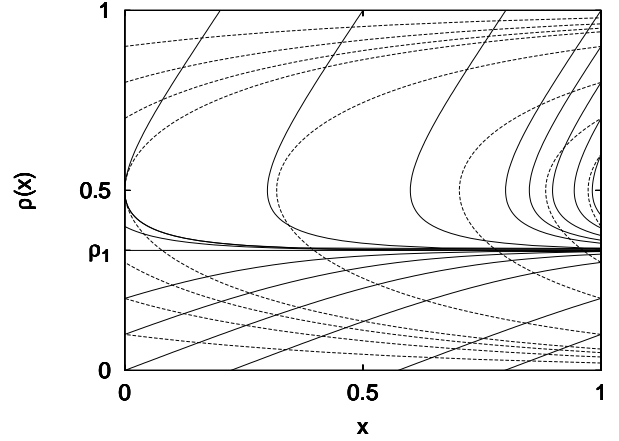}
\caption{\label{ff1} The flow field of the differential equation
  (\ref{diffgen}) for $\Omega_A=2$, $\Omega_B=0.5$ and $J=0$. The solid
  curves represent the complementary-density solutions given in
  eq. (\ref{comp}), whereas dashed curves represent the
  equal-density solutions given in eq. (\ref{equal}).}
\end{figure}

Now, this phenomenon is investigated in detail in the case of asymmetric
lane change. 
Since the current is zero if $\alpha=\beta$, the solutions of 
the hydrodynamic equations are those given in eq. (\ref{equal}) and
eq. (\ref{comp}) (see Fig. \ref{ff1}).
The argumentation about the quasistationarity of the fluctuations of the
global density presented in the case $K=1$ apply to the case $K\neq
1$, as well. 
Thus, in the open system, the density profiles averaged over
configurations with a fixed global density, $\rho_{\varrho}(x)$
and $\pi_{\varrho}(x)$, can be constructed from the solutions 
(\ref{equal}) and (\ref{comp}). 
\begin{figure}[h]
\includegraphics[width=1.0\linewidth]{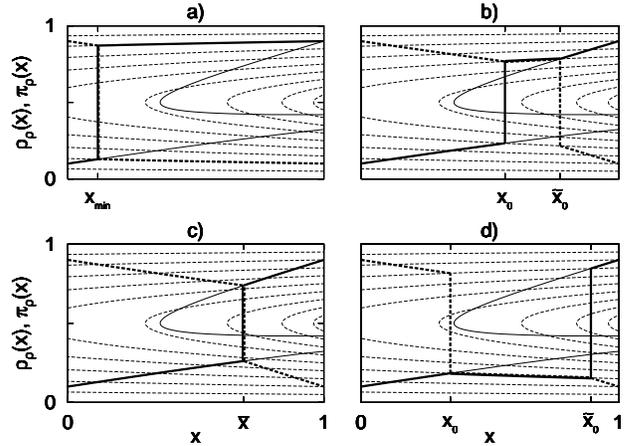}
\caption{\label{ff2} Density profiles in lane $A$ (thick solid lines)
  and in lane $B$ (thick dashed lines) which
belong to a fixed global 
particle density $\varrho$, at four different values of $\varrho$.
The flow field of eq. (\ref{diffgen}) is indicated by thin lines. 
The rates are $\Omega_A=0.5$, $\Omega_B=0.25$ and $\alpha=\beta=0.1$.}
\end{figure}
 
\begin{figure}[h]
\includegraphics[width=0.9\linewidth]{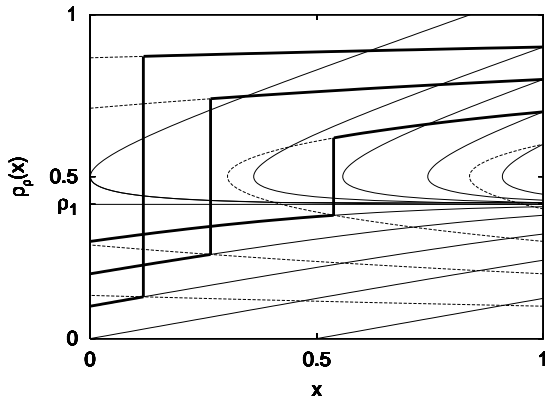}
\caption{\label{ff3} The shock in lane $A$ at the left most possible
  position for $\Omega_A=0.5$, $\Omega_B=0.25$ and for different rates $\alpha$($=\beta$).}
\end{figure}

The structure of these profiles is identical to those obtained in the
case $K=1$ (see Fig. \ref{ff2}). 
Generally, they consist of three segments (see Fig. \ref{ff2}b):  
An equal-density segment is located in the middle part of the system
in the domain $[x_0,\tilde x_0]$. This region is connected with the 
left-hand boundary by a complementary-density segment, which is continuous at
$x=0$, i.e. $\rho_{c,l}(0)=\alpha$,  $\pi_{c,l}(0)=1-\alpha$, and with
the right-hand boundary by another complementary-density segment,
which is continuous at $x=1$, i.e. 
$\rho_{c,r}(1)=1-\alpha$,  $\pi_{c,r}(1)=\alpha$.  
Each lane contains a shock, which are at the opposite ends of the 
equal-density segment (one at $x_0$, the other one at $\tilde x_0$). 
The location of the equal-density segment, as well as $x_0$
and $\tilde x_0$ are determined by the actual global density. 
If $x_0=\tilde x_0\equiv \overline x$, the equal-density segment is lacking and
$\rho_{c,l}(x)$ and $\rho_{c,r}(x)$ are directly connected by a shock
at $\overline x$ (Fig. \ref{ff2}c).  
Since $x_0$ and $\tilde x_0$ are not independent, the shocks move in a
synchronized way, and their motion is confined to the range $[x_{\rm min},1]$.
If one of the shocks is at $x=1$, the other one is at $x_{\rm min}$, thus, 
the lower bound $x_{\rm min}$ is determined by the
equation $\rho_{c,l}(x_{\rm min})=1-\rho_{e}(x_{\rm min})$, where 
$\rho_{e}(x)$ is the equal-density solution which fulfills 
$\rho_{e}(1)=1-\alpha$ (see Fig. \ref{ff2}a).
The lower bound $x_{\rm min}$ is an increasing
function of $\alpha$ (see Fig \ref{ff3}). 
In the limit $\alpha\to 0$, $x_{\rm min}$ tends to zero and if
$\alpha\to\rho_1$, $x_{\rm min}$ tends to 1, thus, at $\alpha=\rho_1$, 
the shock becomes localized at $x=1$ and the system enters the L-H phase. 
At this point, the density
profile is given by the special complementary-density solution:
$\rho_c(x)=\alpha$, $\pi_c(x)=1-\alpha$.

\begin{figure}[h]
\includegraphics[width=0.8\linewidth]{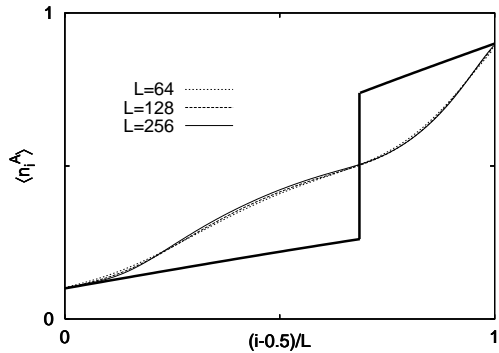}
\caption{\label{coexnum} Density profiles in lane $A$ obtained by numerical
  simulations  with rates  $\alpha=\beta=0.1$, $\Omega_A=0.5$,
  $\Omega_B=0.25$ and for different system sizes (thin lines). The
  thick line represents the density profile $\rho_{\varrho}(x)$
 at global density $\varrho=1/2$ in the continuum limit.}
\end{figure}

Similarly to the case $K=1$, the stochastic variation of the global density
$\varrho(t)$ is described by a bounded symmetric random
walk.  
The stationary density profile $\rho(x)$ can be obtained from 
the profile $\rho_{\varrho}(x)$ at a fixed global
density by averaging it over $\varrho$. 
For $K\neq 1$, we could not carry out the averaging analytically,
nevertheless, we can gain some information on the stationary density
profiles at $\overline x$ without the knowledge of the entire profiles. 
At the point $\overline x$, the relation 
$\rho_{\varrho}(\overline x)=\pi_{\varrho}(\overline x)$ 
holds for all $\varrho$. This, together with the relation $\rho(x)=1-\pi(x)$
following from eq. (\ref{sym2}) implies that 
$\rho(\overline x)=\pi(\overline x)=1/2$.
As it is shown in Appendix B, 
the ratio of the first derivatives of the stationary density profile 
in lane $A$ on the
two sides of the point $\overline x$ is 
\be
\frac{d\rho(\overline x-)}{dx}/\frac{d\rho(\overline x+)}{dx}=
\frac{K+(1-K)\rho_0}{1-(1-K)\rho_0},
\label{rat}
\ee    
where $\rho_0\equiv \rho_{c,\alpha}(\overline x)$.
In the case $K<1$, $\rho_0<\rho_1<1/2$ always holds, 
hence this ratio is smaller
than 1 and the first derivative of the density profile is 
discontinuous at $\overline x$.
Furthermore, it is clear that the stationary density profile
 is identical to the complementary-density solution in the
interval $[0,x_{\rm min}]$, since this domain is forbidden for the shocks.   
We have performed numerical simulations for finite systems 
of size $L=64,128$ and $256$ 
and measured the density profiles in the same way as in the symmetric
case at the coexistence line. 
Results are shown in Fig. \ref{coexnum}. 
 
\section{Discussion}
\label{discussion}

In this work, a two-lane exclusion process was studied,
where the particles are conserved in the bulk and
each lane can be thought of as a totally asymmetric simple
exclusion process with nonconserving kinetics in the bulk. 
As a consequence, the model
unifies the features of the particle conserving and bulk
nonconserving exclusion processes, as far as the dynamics of the
shock is concerned. Namely, it exhibits
phases with a localized shock in one lane, while the other one acts as a
nonhomogeneous bulk reservoir, the position-dependent density
of which is determined by the dynamics itself in a self-organized manner.
On the other hand, the model undergoes a discontinuous phase
transition at the coexistence line, where 
delocalized shocks form in both lanes and move in a synchronized way.
Here, the global density of particles behaves as an 
unbiased random walk, similarly to the TASEP at the
coexistence line, however, the density profiles in the coexisting 
phases are not constant here. 

Although we considered throughout this work exchange rates
proportional to $1/L$, one may imagine other types of scaling. 
For the TASEP with Langmuir kinetics, 
shock localization is observed at the coexistence line when
the creation and annihilation rates vanish proportionally to $1/L^a$ 
with $1\le a<2$ \cite{js}. It might be worth examining 
whether the synchronization of
shocks in the present model persists for lane change rates
vanishing faster than $1/L$.   

In the limit of large lane change rates, $K={\rm const}$,
 $\Omega_A\to\infty$, the density profiles and 
the boundaries of the phases exhibiting
a localized shock are related to the zeros of the source term in
the hydrodynamic equation, which is generally valid for systems with 
weak bulk nonconserving kinetics \cite{frey,pierobon,reichenbach}.
However, different behavior is observed in general when the lane change rates 
$\omega_A$ and $\omega_B$ are finite in the limit $L\to\infty$. 
In this case, the particle current from one lane
to the other may be finite at the boundary layers. 
These currents add to the incoming and/or outgoing currents at the boundaries, 
which may lead to nontrivial bulk densities even in the case $K=1$, 
where the profiles are constant.   

Possible extensions of the present model are obtained when 
different exit and entrance rates or different hop rates 
in the two lanes are taken into account. Nevertheless, these generalized
versions are more difficult
to treat because of the reduced symmetry compared to that of the
present model.
 
\appendix
\section{}
We derive here an approximative expression for the current 
in the H-H phase in the limit of small lane change rates. 
Assuming that $J\ll (\frac{1}{2}-\beta)^2$, 
we may expand the right-hand side of
the differential equation  (\ref{diffgen}) in a Taylor series up to 
first order in $J$. 
Integrating the differential equation obtained in this way yields
\beqn
F(\rho)\equiv \frac{1}{K-1}\ln(\rho-\rho^2)+  
J\left\{\frac{K}{(1-K)^2}\left[\ln\frac{\rho}{1-\rho}-\frac{1}{\rho}\right]+\right.
\nonumber \\
\left.\frac{1}{(1-K)^2}\left[\ln\frac{\rho}{1-\rho}+\frac{1}{1-\rho}\right]
\right\}=\Omega_Ax+{\rm const}, \qquad 
\label{first}
\eeqn
where the first term on the left-hand side is just the equal-density
solution for $J=0$. 
The solution which obeys the boundary condition $\rho(1)=1-\beta$ 
is $F(\rho)-F(1-\beta)=\Omega_A(x-1)$.
From this equation, we get for the density at the left-hand boundary,
$\rho_0\equiv\lim_{x\to 0}\rho(x)$, the implicit equation:    
\be
F(\rho_0)-F(1-\beta)=-\Omega_A.
\label{e1}  
\ee       
On the other hand, we have another relation between $J$ and $\rho_0$:
\be
J=\rho_0(1-\rho_0)-\beta(1-\beta),
\label{e2}
\ee 
thus, we have closed equations for current. 
Note that the leading order term on the left-hand side of
eq. (\ref{e1}), which comes
from the difference of the leading terms of $F(\rho)$ evaluated at
$\rho_0$ and $1-\beta$, is $\mathcal{O}(J)$, while the next-to-leading
contribution in eq. (\ref{e1}) is $\mathcal{O}(J^2)$. 
Expressing $\rho_0$ from eq. (\ref{e2}) and expanding it for small
$J$, we get
$\rho_0=1-\beta-\frac{J}{1-2\beta}-\frac{J^2}{(1-2\beta)^3}+\mathcal{O}(J^3)$.
Substituting this expression into eq. (\ref{e1}) gives an implicit
equation for $J$. Assuming that $J\ll \beta$ and expanding the terms
containing $J$ in this equation in Taylor series, we obtain finally 
\be 
\Theta+\frac{1+K}{2}\Theta^2+\mathcal{O}(\Theta^3)=\frac{\Omega_A}{1-2\beta},
\label{quadr}
\ee
where $\Theta\equiv\frac{J}{(1-2\beta)\beta(1-\beta)(1-K)}$. 
For small $\Theta$, which amounts to $\Omega_A\ll 1-2\beta$, we get a
  good approximation for the current by solving this quadratic equation
 and arrive at eq. (\ref{smallJ}).   

\section{}

As we have seen, the positions of the shocks in lane $A$ and lane $B$ are not
independent, thus, we may define thereby a function $\tilde x_0(x_0)$,
which is given implicitly by the equation 
$\rho_{e}(\tilde x_0)=1-\rho_{c,r}(\tilde
x_0)$, where $\rho_e(x)$ is the equal-density solution which satisfies 
the condition $\rho_{c,l}(x_0)=\rho_e(x_0)$.
In the following, we shall denote the density in lane $A$ at $x_0$, if  
the shock is located at $x_s$, by $\rho_{x_s}(x_0)$.
First, we notice that at the reference point $x_0$
($x_0<\overline x$), $\rho_{x_s}(x_0)$=$\rho_{c,l}(x_0)$ holds
whenever the shock in lane $A$
resides between $x_0$ and $\tilde x_0(x_0)$, i.e. $x_0<x_s<\tilde
x_0(x_0)$. Similarly, $\rho_{x_s}(\tilde x_0(x_0))$=$\rho_{c,r}(\tilde
x_0)$ if $x_0<x_s<\tilde x_0(x_0)$.
When the shock is outside this interval 
($x_s\notin [x_0,\tilde x_0]$), 
then  $\rho_{x_s}(x_0)$=$\rho_{e}(x_0)$, where $\rho_{e}(x)$ is
some equal-density solution determined by the global density   
and $\rho_{x_s}(x_0)>1/2$ or $\rho_{x_s}(x_0)<1/2$ if
$x_s<x_0$ or $x_s>\tilde x_0(x_0)$, respectively.
As a consequence of the particle-hole symmetry, 
the relation $\rho_{x_s}(x_0)=1-\rho_{\tilde
  x_s(x_s)}(x_0)$ holds if $x_s\notin [x_0,\tilde x_0]$.  
Furthermore, in the stationary state, the probability that $x_s<x_0$ is 
equal to the probability that $x_s>\tilde x_0(x_0)$ 
for any $x_0\le \overline x$. 
Therefore the contribution to
the average profile is $1/2$ when $x_s\notin [x_0,\tilde x_0]$.  
We can thus write for the average densities at $x_0<\overline x$ and
$\tilde x_0(x_0)>\overline x$
\beqn 
\rho(x_0)=p(x_0)\rho_{c,l}(x_0)+[1-p(x_0)]\frac{1}{2}, \nonumber \\
\rho(\tilde x_0)=p(x_0(\tilde x_0))\rho_{c,r}(\tilde
x_0)+[1-p(x_0(\tilde x_0))]\frac{1}{2},
\label{average}
\eeqn
respectively, where $p(x_0)$ is the probability that the shock in lane
$A$ resides in the interval $[x_0,\tilde x_0(x_0)]$, and 
$x_0(\tilde x_0)$ is the inverse function of $\tilde x_0(x_0)$.
For the spatial derivatives of the densities at $x_0$ and $\tilde
x_0(x_0)$, we get   
\beqn
\rho'(x_0)=p'(x_0)\left(\rho_{c,l}(x_0)-\frac{1}{2}\right)+p(x_0)\rho_{c,l}'(x_0),
\nonumber \\
\rho'(\tilde x_0)=p'(x_0)x_0'(\tilde x_0) \left(\rho_{c,l}(\tilde
x_0)-\frac{1}{2}\right)+p(x_0(\tilde x_0))\rho_{c,r}'(\tilde x_0), \nonumber
\\
\eeqn 
where the prime denotes derivation.
Using that $p(\overline x)=0$ and $\rho_{c,l}(\overline
x)=1-\rho_{c,r}(\overline x)$, we obtain for the ratio of the left-
and right-hand side
derivatives at $\overline x$:
\be
\rho'(\overline x-)/\rho'(\overline x+)=|\tilde x_0'(\overline x)|.
\ee 
Expanding the functions $\rho_{c,l}(x)$, $\rho_{c,r}(x)$ and
$\rho_{e}(x)$ in Taylor series up to first order in $x$ 
around $x=\overline x$, we obtain 
\be
|\tilde x_0'(\overline x)|=
\frac{\rho'_{c,l}(\overline x)-\rho'_{e}(\overline x)}
{\rho'_{e}(\overline x)-\rho'_{c,r}(\overline x)}. 
\ee
Using eq. (\ref{diff1}), these derivatives can be given 
in terms of $\rho_0\equiv \rho_{c,\alpha}(\overline x)$
and we arrive at eq. (\ref{rat}).

\acknowledgments
The author thanks L. Santen and F. Igl\'oi for useful discussions.
This work has been supported by the National Office of Research and 
Technology under grant No. ASEP1111 and by the Deutsche
Forschungsgemeinschaft under grant No. SA864/2-2.


\end{document}